**Effect of the crystal size on the X-ray diffraction patterns of isolated orthorhombic starches: A-type**

Mario E. Rodriguez-Garcia[a,*], Sandra M. Londoño-Restrepo[b], Cristian F. Ramirez-Gutierrez[b], Beatriz Millan-Malo[a]

[a]*Departamento de Nanotecnología, Centro de Física Aplicada y Tecnología Avanzada, Universidad Nacional Autónoma de México, Campus Juriquilla, Querétaro, Qro., C.P 76230. México.*

[b]*Posgrado en Ciencia e Ingeniería de Materiales, Centro de Física Aplica y Tecnología Avanzada, Universidad Nacional Autónoma de México, Campus Juriquilla, Querétaro, Qro., C.P 76230. México.*

[*]**Corresponding author**: Mario E. Rodríguez-García, Departamento de Nanotecnología, CFATA, UNAM, Querétaro, Qro, México. E-mail: *marioga@fata.unam.mx,* Fax: (+52-442-2381165

**Abstract:**

The x-ray diffraction (XRD) patterns reported for starch in the literature describes it as a semicrystalline polymer that indicates that amorphous and crystalline regions form it, and this is commonly accepted. However, these patterns have not been well interpreted according to the XRD theory, and the crystal size effect has been neglected. A simulation of the effect of the crystal size on the shape and width on XRD patterns for A-type (orthorhombic structure, ICDD card No. 00-043-1858) using PDF-4 software shows that the patterns of starches are governed by the elastic and inelastic scattering producing wide peaks. For isolated starches XRD patterns,  the separation of inelastic and elastic scattering is not possible, then the calculation of the crystalline percent or crystalline quality does not make any physics sense.

1. **Introduction**

Starch is a "microparticle" mainly formed by two macromolecules called amylose and amylopectin, being one of the most important carbohydrates in nature. Different starch sources are coming from tubers, fruits, cereals, seeds, legumes, among others. The uses of native and modified starches are directly related to their physicochemical properties. One of these properties is the crystalline structure, but the literature does not provide answers to many questions regarding starch architecture or crystalline structure for amylose and amylopectin (Tester, Karkalas, & Qi, 2004). Therefore, it is still an open problem from a physics point of view with strong influence in the food industry.

All crystalline materials in nature are structurally classified by the form where the atoms are arranged in a lattice which is a mathematical abstraction called Bravais lattices (see Fig. 1) (Kitel, 1996). For each one of these materials, the minimal number of atoms that are repeated in each point of the lattice forms the atomic base. The atomic base and the crystal is formed by three-dimensional translations of the atomic base in every site of Bravais lattice.

It is well known that some materials present more than one crystalline structure, as it is the case of carbon that forms three-dimensional polymorphic structures such as graphite with hexagonal structure, the diamond with face center cubic (FCC) structure, and graphene which is a two-dimensional structure. Polymorphism is defined as the ability of any material to be in more than one crystalline structure. It can be found in many crystalline material that include minerals, semiconductors, and metals, and it is relevant in some fields such as food, pharmaceutical, pigments, and agrochemical due to their physicochemical properties depend strongly on the crystalline structure.

Powder X-ray diffraction (XRD) techniques are normally used to study nanosized systems because these are nondestructive techniques. However, the conventional XRD diffraction technique in Bragg Brentano configuration has some limitations that complicate the analysis, as it is the case of soft matter and organic nanoparticles. Soft matter is a variety of physical systems such as liquids, polymers, colloids, granular materials, and biological systems, where the interpretation and analysis of XRD patters that involves elastic and inelastic scattering are not adequately considered. In fact, it is an emerging study area (Stribeck, 2002).

In these materials, the length distribution of the molecular chain (molecular mass distribution) and the crystal size distribution or heterogeneity of structural entities like fibrils or lamellar stacks, play a very important role to the XRD pattern. It is usual to confuse the diffracted peaks produced by a sample that has low crystalline quality with peaks from a nanocrystalline material, as in both cases, samples produce broad diffracted peaks, as is reported for isolated starches from different botanical sources (Stribeck, 2002).

Regarding polymorphisms in starch, four types have been reported: A, B, C, and V type. A-type, that is mainly present in cereals; B-type for starches from most tubers and roots that according to Zobel (1988), crystallize with "broad" and "weak" peaks located at about 5.5 and 17.2° in 2θ scale. While, C-type is considered a superposition of the A and B types, that form a physics and crystallographic point of view do not make any sense.

Finally, V-type is characteristic for amylose complexed with fatty acids and monoglycerides, which took place after starch gelatinization, thus, this type hardly ever appears in native starches. Nonetheless, it has not even reported at least three diffracted peaks to identify this complex or a powder diffraction file.

There is a considerable amount of information in the literature that contradict one another; until know, it has been proposed many crystalline systems for starches, but the existence of these polymorphisms has not been proved.

Imberty et al. (1988) proposed a three dimensional structure for B-type starch with a unit cell that contains 12 glucose residues and parallel-stranded double helices packed where 36 water molecules are located between them. They reported that amylose chains crystallize in a hexagonal lattice that belongs to the P6 space group, with lattice parameters a = b = 1.85 nm and c = 1.04 nm. Nevertheless, the structure reported for amylose in the Powder Diffraction File (PDF) emitted by the International Center for Diffraction Data (ICDD) with card No. 00-043-1858, is orthorhombic I with the following lattice parameters a=10.69, b=11.72, and c= 17.71 nm and is associated with A-type. Although the information to develop this card was described by Imberty et al. (1988), they reported that the amount of diffraction data is not enough to attempt a structure elucidation through conventional crystallographic methods.

Imberty and Perez (1988) pointed out the attempts to get an unambiguous understanding of the crystalline arrangement have not been established due to the inadequate amount of data, the small crystalline size that is in fact in the nanometric scale, "the imperfection of the crystallites", and the large unit cell. Regarding C-type, the said finding does not make a physics sense from a crystallographic point of view because any structure must have one of the 14 Bravais lattices. Fig. 1 shows the Bravais lattices and describes the crystalline systems, showing the main characteristics of lattice parameters and their angles. These lattices grouped into seven lattice systems: triclinic, monoclinic, orthorhombic, tetragonal, rhombohedral, hexagonal, and cubic.

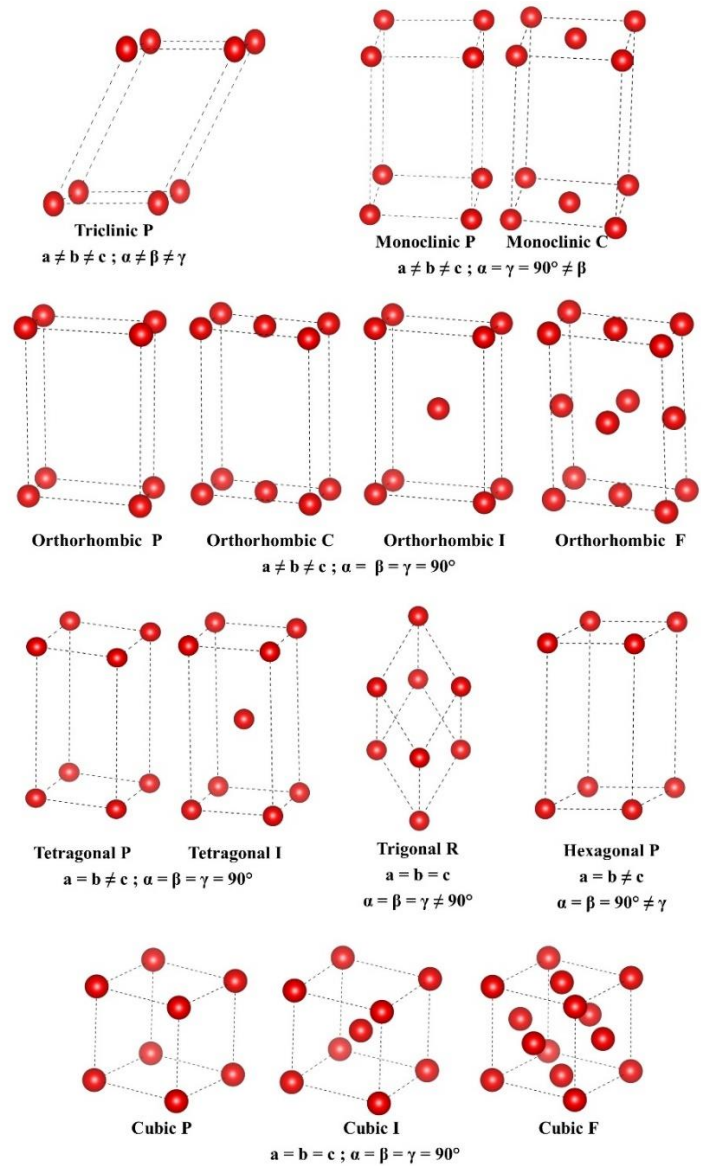

**Figure 1.** Bravais lattice for crystalline structures in nature.

On the other hand, according to Gernat et al. (1993), when two materials are combined, a chemical reaction can or cannot take place; when a XRD pattern is taken, it can change as follows. If there is a chemical reaction, two results are expected: the formation of an amorphous material or a new crystalline phase or phases. Thus, some changes can be seen in the diffraction pattern concerning the initial samples before any chemical reaction: in the first case, the amorphous material will contribute to the background, but in the second case, new peaks will be found. Recently, Londoño-Restrepo et al 2019 showed that for nanometric crystals of hydroxyapatite from bovine bone, the X-ray pattern is formed by elastic and inelastic contributions originating broad peaks. Then, broad peaks do not have any relationship with the crystalline quality of the sample at nanometric crystal scales. In fact, the TEM images of the material show the existence of ordered structures.

It is important to point out that any of the fourteen Bravais lattices must correspond to the new crystalline material. However, when the chemical reaction does not take place, the XRD pattern is a superposition of every single pattern from the initial materials since any new phase is not present. Then, if it is reported as a new structure, as it is the case of C-type starch, it does not make physical sense.

According to Imberty et al. (1991), the difference between A and B-types arises from the water content and the way these pairs are packed in the respective crystals. On the other hand, these authors reported that native starch granules produce diffraction patterns with low crystalline quality and crystallinity percent from 15 to 45%. But this is a misinterpretation of the diffracted patterns since these are obtained from nanocrystals.

The primary approach to identify a crystalline structure is using a PDF. The identification is obtained by comparison of the diffraction curve of the studied sample (diffracted peaks) and a known standard or a database in which the allowed diffracted peaks and their Miller indices, as well as the experimental and calculated intensities, are reported. Notwithstanding, starches samples have been identified by using a PDF (ICDD card No. 00-043-1858) that corresponds to orthorhombic structure. At this point, it is important to recall that the peak located at about

5.64 in the 2θ scale used to identify the B-type of starch and corresponds to the (100) direction for the hexagonal structure.

According to Wennerström (2014), scattering and diffraction are generally described separately, but from an XRD point of view, they are so closely related that in a conceptual description, they are considered as basically the same phenomenon. The difference is in the long-range order exhibits by the sample for diffraction.

The objective of this work is to propose an interpretation of the XRD patters of starches including the elastic and inelastic scattering contributions. The proposed physical explanation considers the effect of the nanometric size of the crystals into these starches on the so XRD pattern of amorphous biopolymers. The concept of crystalline percent calculated from the whole XRD diffraction patterns is revised. Finally, the PDF-4 software is used to simulate and study the effect of the crystal size on the shape and width of the diffraction patterns, although this program does not consider the amorphous contribution.

**2.. Materials and methods**

**2.1. Starch sample**

Corn starch rich in amylopectin (pcode: 100952946) and corn starch rich in amylose (lot: 64H1025, A-7043) from Sigma-Aldrich (Germany) were studied in this work.

**2.2. The nanocrystalline character of starch**

It is well known that any starch is composed of nanocrystals. Haaj et al. (2016) studied starch nanocrystals obtained through ultrasound (particle size of 38 nm), as well as after an acid hydrolysis treatment (58 nm particle size), starting from waxy maize starch. They reported that native waxy-maize exhibits the A-type structure and a crystalline index (CrI) around

47%. The X-ray diffraction patterns for the samples show broad peaks and high similitude. It is well established that nanoparticles produce X-ray diffraction patterns with broad peaks due the scattering of the nanocrystals. Here, it is important to emphasize that there is a difference between crystalline starch nanoparticles (CSN) that are present in native starch and starch-based nanomaterials that are obtained by different crystallization or re-crystallization processes. Gonga et al. 2016 showed that the starch nanocrystals (5–10 nm) obtained by an acid hydrolysis exhibit a similar crystalline pattern to the maize starch granules classified as A-type starch these patterns showed broad peaks. This fact supports the contention that the crystal structure of the as-prepared CSN is inherited from the crystallinity of the normal corn starch granules that is characteristic of monoclinic unit cells. It means that the native and isolated starches are composed of nanocrystals and that the interpretation of the XRD diffraction patterns must be analyzed in detail.

**2.3. The crystalline percent calculation through the X-ray diffraction pattern**

Along time, different researchers have published the quantification of the crystalline percent or crystalline index for different isolated starches from several sources. However, in this section, the focus is centered on some representative works to clarify some misunderstanding about the possibility to obtain information about this parameter through an XRD pattern. Gonga et al. (2016) reported the crystalline percentage for starch nanocrystals and starch nanocrystal obtained by the acid method from 33.4% to 35.2%. However, if the XRD pattern corresponds to nanoparticles, the shape and broad peaks are governed by the X-ray elastic and inelastic scattering, it means that the pattern does not correspond to pure elastic scattering because it is also formed by the inelastic scattering signal. Then, the calculation of the crystalline percent or crystalline index does not make any physics sense. In fact, these authors showed TEM images in which it is clear the existence of high ordered regions that have high crystalline order and some amorphous areas formed by the sintering acid process. Amini and Ali Razavi (2016) used the same procedure to calculate the crystalline quality of starch nanocrystals, and they found values from 28.1 to 36.6%, but again, these XRD patterns correspond to nanocrystals. They also showed TEM images, but the calculation of the crystalline quality was not done.

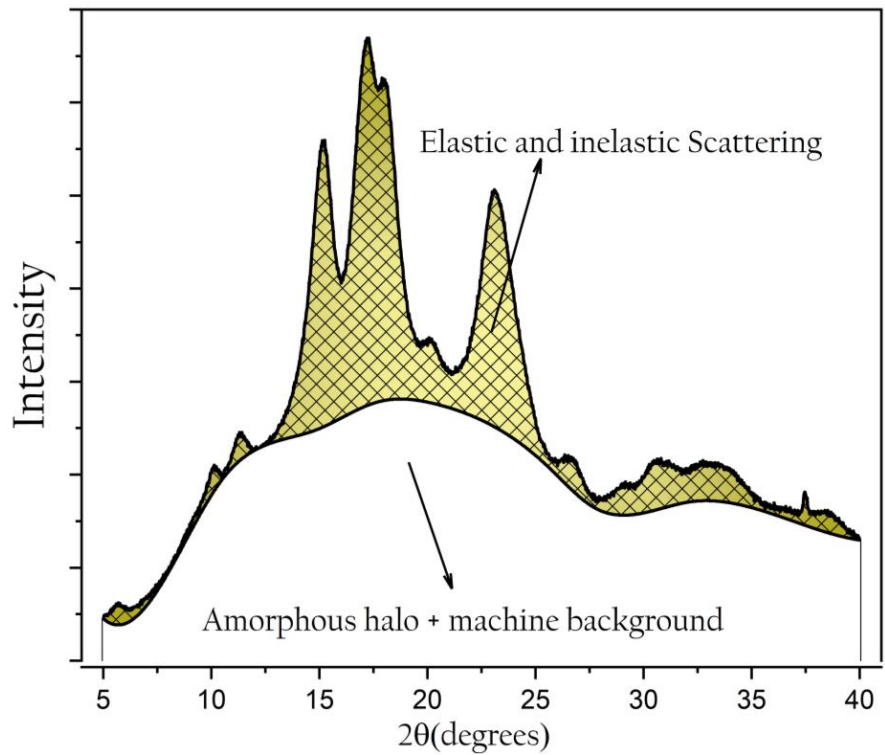

**Figure 2**. Characteristic X-ray diffraction pattern of corn starch rich in amylopectin (orthorhombic structure) in which elastic and inelastic scattering, amorphous region and machine background are identified.

Fig. 2 shows a characteristic XRD on pattern of corn rich in amylopectin (orthorhombic structure). Usually, three regions are identified in a pattern: crystalline, amorphous, and noise (machine background). However, it is important to show that the region that has been used to describe the crystalline contribution of a nanocrystalline sample, is formed by elastic and inelastic contributions; this fact does not allow to determine each one of these signals independently. In fact, there is in the literature a paper called: "Separating diffraction from scattering: the million-dollar challenge" by (Laven, 2010). This is an instrumental and physics problem in different areas of physics in which both inelastic and elastic scattering are present. Finally, according to Fig. 2, the regions that can be defined in a XRD pattern of a nanocrystal are: the elastic and inelastic scattering region that involves information regarding the nano-size character of the crystals and the crystalline quality, the amorphous region, and the background.

## 2.4. Theoretical framework: X ray diffraction in food science

The diffraction is a phenomenon defined as spatial energy redistribution because of the interaction of the electromagnetic wave with the matter where the space between its atoms is close to the incident wavelength. The geometrical configuration of maximum and minimums of the intensity of the radiation constitute an X-ray diffraction pattern. Therefore, the energy redistribution is associated with the geometry of the object, and it defines the pattern, which is the case of the X-ray radiation and the atomic arrangement of the matter (Kitel, 1996; Cullity & Stock, 2014). The X-ray diffraction by matter is a consequence of a combination of two phenomena. The first one is related to the scattering of individual atoms, and the second one is the interference between the waves scattered by these atoms.

Nowadays, the most common diffraction method is the powder diffraction because it is a practical method that can be used when a single-crystal sample is not available, as is the case of the more often starchy materials. In this case, the sample is reduced to a fine powder, each particle would be a crystal or assembly of smaller crystals, and they are randomly oriented concerning the optical axis. The result is that every set of planes will be able to diffract. The relation between the peaks positions, intensity, and shape of the diffraction pattern is a mixing of the position of every single atom, as well as the length of lattice parameters. It means that any calculation of the intensity gets started with structure factor (Kitel, 1996; Cullity, & Stock, 2014) because it contains information about the atomic base of the crystal and the electronic density. In the case of starch material, it is still an issue as was mentioned above. Nonetheless, other factors affect the relative intensity of the diffraction peaks in the case of a powdered sample. These factors are polarization factor, structure factor, multiplicity factor, Lorenz factor, absorption factor, and temperature factor (Cullity, & Stock, 2014). Moreover, the diffraction peaks are modified by defects such as dislocations, vacancies, and atomic substitutions. The "size" of the crystal and strain are other parameters that alter the shape and width of the diffraction peaks (Cullity, & Stock, 2014).

Furthermore, the scattering experiments are carried out in a variety of angular regions that are typically identified by the distance (R) between the sample and the detector. Classical powder X-ray diffraction and scattering are run in the subarea of wide angle X-ray scattering (WAXS) where the parameter R is between 0.05-0.2 m, and it is focused on study the arrangement of chain segments in the case of the soft matter. Small-angle X-ray scattering (SAXS) is also a conventional configuration to investigate the soft matter.

## 2.5. X-ray diffraction characterization

X-ray diffraction patterns of the isolated starches were carried out with a Rigaku-Ultima IV diffractometer. The operating conditions were 40 kV and 30 mA, CuK$_\alpha$ radiation wavelength of λ=0.1540 nm, from 5 to 35° in a 2θ scale with a step size of 0.2°/min to obtain patterns with low noise. The powder samples were densely packed into an aluminum holder. X-ray diffraction patterns of powder samples of corn rich in amylose and corn rich in amylopectin from Sigma Aldrich were carried out as references.

**2.6. PDF-4 software to study the effect of the crystal size**

The effect of the crystal size on the X-ray patterns has been studied in detail by using the PDF-4 software. This software considers the peak shape analysis to provide information on crystal size and strain distributions of samples, as well as the Bragg intensities that gives information about possible preferred orientation and texture effects in powdered samples (Needham & Faber, 2003; Kabekkodu, Faber, & Fawcett, 2002). On the other hand, a powder diffraction files (PDF) is required for the analysis of the effect of the crystal size on the X-ray diffraction pattern (Faber & Fawcett, 2002). The PDF-4 software requires the structural information of the material for the simulation. This information is provided for the PDF card or a Crystallographic Information File (CIF).

**3. Results**

**3.1. Size effect on the X-ray diffraction patterns**

In the case of diffraction in nanocrystals, the crystal size governs the diffraction pattern, originating broad peaks. The effect of the crystal size has been studied using PDF-4 and assuming that there is no changes in the crystal structure.

Fig. 3 shows the X-ray diffraction patterns of corn starch rich in amylose and amylopectin, as well as the simulated patterns for the orthorhombic structure considering the JCPDS card No. 43-1858. In this simulation, the crystal size effect on the X-ray diffraction pattern for the orthorhombic structure is simulated by changing the crystal size from 2 to 20 nm. The simulation for the hexagonal structure so far cannot be carried out due to that it is necessary to have the Powder diffraction file.

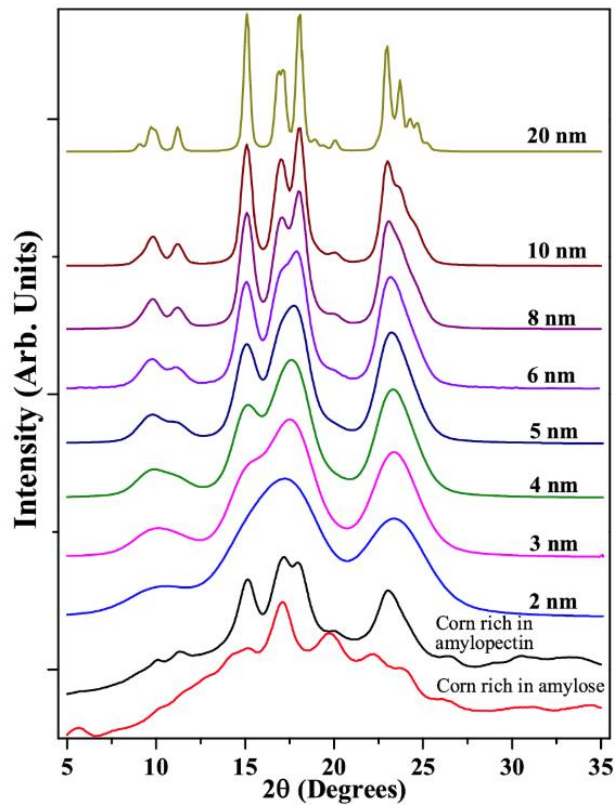

**Figure 3.** X-ray diffraction patterns of starch rich in amylose, starch rich in amylopectin, as well as the simulated patterns for orthorhombic structure changing the crystal size for 2 to 20 nm.

As can be seen in Fig. 3, the simulates patterns exhibit broad peaks for the smallest crystals due to that the main factor is the simultaneous contributions of the elastic and inelastic X-ray scattering, this is because the crystal size of the sample is close to the incident wavelength; in this case, the $CuK_\alpha$ radiation of the system. This simulation does not consider the amorphous contribution of the material.

If the crystal size increase from 10 to 20 nm in the simulation, the broad peaks become defined due to the X-ray inelastic scattering decreases and the resulting diffracted peaks are the result of the elastic contribution to the X-ray diffraction in the orthorhombic crystalline structure.

Here, it is evident that the crystal size plays an important role in the shape and width of the XRD patterns. This simulation clearly shows that the reported XRD diffraction patterns for isolated starches (A-Type), in fact correspond to nanocrystals, and it does not represent poly-crystals with low crystallinity as it has been reported in the literature. These broad peaks are related to the existence of nanocrystals within the starch granules.

In the case of crystals with size domain of micrometers, the XRD patterns can thin and well defined (assuming good crystalline quality) while in the case of the diffraction in nanocrystals (some multiples of the X-ray wavelength used), where X-ray inelastic scattering governs the shape and width of the peaks, these patterns are composed of broad and overlapped peaks as was shown in the XRD patterns simulated for the orthorhombic structure (see Fig. 3).

The technique used to obtain these patterns was powder diffraction; it means that all possible diffractions planes must be present. Patterns for amylose and amylopectin exhibit broad peaks, that do not necessary represent crystals with low crystalline quality.

### 4. Conclusions

The study of the crystalline structure of starches from different botanical sources by XRD diffraction is still an open issue according to the reported in the literature.

The simulation of the effect of the crystal size on the shape and width for the XRD patterns in starches, showed clearly that these patterns are governed by elastic and inelastic scattering originated by the nanocrystal size and by the amorphous contribution of the amylose and/or amylopectin fractions.

The XRD diffraction patterns deconvolution, peaks indexation, and baseline corrections used for different authors to characterize the type of starch do not make any physic sense because the process to calculate these parameter does not consider the inelastic scattering contribution coming from the crystal size that is close to X-ray wavelength.

Considering the discussion, in the case of XRD patterns for isolated starches, they identified nanocrystals, and then the calculation of the crystalline percent and crystalline quality cannot be done. The crystalline quality can be obtained by an in-depth analysis of the TEM images in the Fourier space.

The polymorphism, as well as the crystalline structures in starches are still an open question that requires a deep analysis to avoid any misunderstanding in the starch structure classification. It is necessary to define the crystalline structure taking in to account the Bravais lattices.

**Acknowledgments**

Sandra M. Londoño-Restrepo and Cristian F. Ramirez-Gutierrez thank Consejo Nacional de Ciencía y Tecnología-México (CONACYT) for their Ph.D scholarship. Authors are eternally grateful to Laboratorio Nacional de Caracterización de Materiales (LaNCaM)-CFATA for characterization material support.